\documentclass{rsauthor}
\usepackage[numbers]{natbib}
\usepackage{graphicx,subfig}
\DeclareGraphicsExtensions{.pdf}
\def\vec#1{{\bf#1}}

\def\proj#1{|#1\rangle\langle#1|}
\def\one{\mathbb{I}}
\def\ket#1{| #1 \rangle}
\def\bra#1{\langle #1 |}

\def\ave#1{\langle #1 \rangle}

\def\diag{\operatorname{diag}}

\def\eps{\epsilon}

\jname{Phil. Trans. A}

\begin{document}

\title{Quantum system characterization with limited resources}

\author{D. K. L. Oi$^a$, S. G. Schirmer$^b$} \address{$^b$SUPA,
  Department of Physics, University of Strathclyde, Glasgow, G4 0NG,
  United Kingdom, $^b$Department of Physics, Swansea University,
  Singleton Park, Swansea, SA2 8PP, United Kingdom}

\date{\today}

\abstract{The construction and operation of large scale quantum
  information devices presents a grand challenge.  A major issue is
  the effective control of coherent evolution, which requires accurate
  knowledge of the system dynamics that may vary from device to
  device.  We review strategies for obtaining such knowledge from
  minimal initial resources and in an efficient manner, and apply
  these to the problem of characterization of a qubit embedded into a
  larger state manifold, made tractable by exploiting prior structural
  knowledge. We also investigate adaptive sampling for
  estimation of multiple parameters.}

\keywords{quantum, control, estimation}


\maketitle

\section{Introduction}

Recently, much effort has been put into the construction of large
scale quantum devices operating in the coherent regime. This has been
spurred by the possibilities offered by quantum communication and
information processing, from secure transmission, simulation of
quantum dynamics, to the solution of currently intractable
mathematical problems~\cite{NatureReview2008}.  Many different physical
systems have been proposed as the basic architecture upon which to
construct quantum devices, ranging from atoms, ions, photos, quantum
dots and superconductors. For large scale commercial application, it
is likely that this will involve scalable engineered and constructed
devices with tailored dynamics requiring precision control.

Due to inevitable manufacturing tolerances and variation, each device
will display different behaviour even though they may be nominally
``identical''. For their operation, they will need to be characterized
as to their basic properties, response to control fields, and noise or
decoherence~\cite{BIRS2007}.  We may also need to know how ideal is the
system in the first place, for instance the effective Hilbert space;
what we may assume to be a qubit may have dynamics involving more than
two effective levels. Extracting this information efficiently and
robustly is crucial.

In a laboratory setting, an experimentalist may have access to many
tools with which to study a system, e.g. spectroscopy and external
probes.  In a production setting, provision of these extra resources may
be difficult, expensive, or impossible to integrate with the device. It
is therefore an important to understand what sort of characterisation
can be performed simply using what is available \emph{in situ}. Ideally,
we would also like to be able to characterize the performance of a
device with as little prior knowledge of its behaviour, e.g. how it
responds to control fields, if this is the information which we are
trying to obtain in the first place.  Characterization using only the
\emph{in situ} resources of state preparation and measurement, even
where it is possible, is challenging, due to the increasing complexity
of the signals, number of parameters to be estimated and the complexity
of reconstructing a valid Hamiltonian from the resulting signal
parameters.  Robust and efficient methods of data gathering and
analysis, preferably as automated as possible, are therefore
essential.  Here, we review progress in tackling these problems and we
present new results for the problem of characterizing the Hamiltonian of
a qubit embedded in a larger state manifold.

\section{System identification paradigms}

A standard tool for determining discrete quantum dynamics is
\emph{quantum process tomography} (QPT). Assume a completely positive
trace preserving map $\Lambda$ acting on a system state $\rho$ by
\begin{equation}
  \Lambda(\rho)=\sum_j \lambda_j \rho \lambda_j^\dagger
\end{equation}
where $\lambda_j$ are the Kraus operators satisfying $\sum
\lambda_j^\dagger \lambda_j = \one$.  QPT is a method by which we can
determine $\{\lambda_j\}$ by seeing how initial states of the system
evolve under the map. For a complete characterisation, a complete set of
initial states $\{\rho_j\}$ and quantum state tomography on the final
states are both required~\footnote{There are variants of QPT which use
ancillas, entangling or other coherent operations which may offer
certain advantages~\cite{MRL2008,SBLP2011}. However they all require
operational resources beyond what we assume for system
characterisation.}.  
In general for a d-dimensional system, we need to use $d^2-1$ input
states, and quantum tomography either requires a measurement with at
least $d^2$ outcomes, for instance, symmetric informationally complete
positive operator-valued measures SIC-POVMs~\cite{SICPOVM}, or
projective measurements in several different bases.

The ability to generate a complete set of initial states is a strong
assumption. In many physical systems, it is only possible to directly
prepare a set of orthogonal states, e.g. by projective measurement, or
even only a single ``fiducial'' state. The usual assumption that we can
generate any state by coherent control of a fiducial state is invalid
when we are trying to determine how to control the system in the first
place. Equally, most systems can only be measured directly in a single
fixed basis, and other measurement bases are assumed to be available
through coherent control.  This shows that QPT cannot be used as a
starting for quantum system identification in a setting where control
has not been already established.  We require a mechanism by which we
can bootstrap our knowledge and abilities until control can be enacted
upon the system.

Assuming the system dynamics can be approximated to a reasonable degree
by Hamiltonian dynamics, the first core challenge is the identification
of the \emph{intrinsic system Hamiltonian} $H=\sum_j E_j \proj{E_j}$,
which can be specified by the energy eigenstates and eigenvalues (up to
rescaling).  Optimum protocols for identification of the Hamiltonian
dynamics depend on the available resources.  One general paradigm
introduced in \cite{SKO2004} assumes a situation where we are restricted
to preparation and measurement in a single fixed basis $\{\ket{e_j}\}$,
i.e., we can prepare the system in one of the basis states $\ket{e_j}$,
allow it to evolve for some time $t$ before projectively measuring in
the same basis, obtaining state $\ket{e_k}$ with some probability
$p_{jk}(t)$.  The computational basis in this case can be defined in
terms of preparation and measurement and our task is to characterize the
system Hamiltonian in this basis, i.e. obtain the eigenenergies $E_j$
and the eigenstates $\ket{E_j}$ (up to a global phase), solely from the
data traces of $p_{jk}(t)$.

Hamiltonian characterization in this paradigm has been considered in a
number of papers.  It has been shown that a generic Hamiltonian for a
single qubit can be recovered from preparation and measurement in a
fixed basis up to a certain set of phases and an unobservable global
energy shift.  The extra phases become relevant only when additional
resources are available that allow us to initialize the system in
non-measurement basis states, or apply control that alters the system
Hamiltonian. In the latter setting is was also shown how the relative
phases for two Hamiltonians could be recovered by composite rotations,
vaguely reminiscent of Ramsey spectroscopy~\cite{Ramsey1949}.  If the
Hamiltonians do not commute or coincide with the preparation/measurement
basis, full control over a single qubit can be realized and the system
and control Hamiltonians can be characterized up to a sign
factor~\cite{SOKC2004}.  The results can be extended to multi-qubit or
higher level systems~\cite{SO2009}.

Upon characterisation of the intrinsic system Hamiltonian the effect
of applying controls must be identified.  We can model this by
assuming a Hamiltonian $H(\vec{\lambda})$ that depends on classical
control field parameters $\vec{\lambda}$.  In the simplest case we may
approximate the response of the system by $H(\vec{\lambda})=H_0+\sum_j
\lambda_j H_j$, where $H_0$ is the intrinsic system Hamiltonian and
$H_j$ are perturbations resulting from the application of control
$\lambda_j$.  This has been considered in~\cite{SKO2004,SOKC2004}
where the effect of multiple control fields on a single qubit were
characterized with respect to a reference Hamiltonian.  For coherent
operation, \emph{incoherent effects} should be small but they
will still need to be characterized. Although complete
characterization of the dynamics for open systems is a daunting task
under certain simplifying assumptions on the type of decoherence,
e.g., pure dephasing or relaxation in a natural basis, the number of
parameters to be estimated can be reduced and the relevant information
extracted~\cite{OiSchirmer2011,CGOSWH2006,SO2009A,qph1012_4593}.

Although the general characterization paradigm described is quite
restrictive, in some cases further restrictions must be imposed to
deal with limited resources.  Characterization of the coupling
constants in a spin network when only a subset of the spins at the
boundary can be measured and initialized is an
example~\cite{BMN2008,BMN2011,FMK2011}. Another is estimation of
leakage out of a subspace or coupling to unknown states, where we
generally cannot measure the populations of these states directly.
General bounds on subspace leakage were derived in~\cite{DSOCH2007}
based on the Fourier spectrum of the observed Rabi oscillations.  This
approach is useful when there are potentially many states very weakly
coupled to the subspace of interest.  In many cases, however, leakage
may be due to coupling to a small number of states outside the
subspace, e.g., when we encode a qubit using the lowest two states of
a slightly anharmonic oscillator.  In this case a control applied to
the qubit transition will induce some coupling to the third (nuisance)
level, giving rise to unwanted dynamics.  Characterization of this
coupling allows the design of pulses that can suppress such unwanted
excitations.

\section{Hamiltonian estimation for embedded qubit}

Formally, we consider a three-level system subject to a control field
resonant with the $1-2$ transition (see also~\cite{Leghtas2009}), where
$\ket{1}$ and $\ket{2}$ can be regarded as the qubit states.  Assuming a
constant amplitude field $f(t)=A \cos(\omega t)$, transforming to a
rotating frame and making the rotating wave approximation, this leads to
an effective Hamiltonian of the form
\begin{equation}
 \label{eq:H1}
  H= \begin{pmatrix}
        0 & d_1 & d_3 \\
      d_1 &   0 & d_2 \\
      d_3 & d_2 & \delta
     \end{pmatrix}.
\end{equation}
We choose a rotating frame where the off-diagonal
elements $d_n$ are real and positive.  If higher order
processes such as two-photon transitions between states $1$ and $3$ are
negligible, we can assume $d_3=0$.  The objective is to characterize
both the qubit transition coupling $d_1$ as well as the coupling to the
nuisance level $d_2$ and the detuning $\delta$ as a result of the
anharmonicity.

If the system can be initialized in the basis states $\ket{n}$ for
$n=1,2,3$ and we perform complete projective measurements in the same
basis, then the probabilities
\begin{equation}
  p_{k\ell} = |\bra{k} e^{-i H t} \ket{\ell}|^2
\end{equation}
can be determined for all $k,\ell$ for different times $t_j$.  However,
if we can only initialize and measure the system in the ground state
then only a single population evolution trace $p_{11}(t)$ is available.
For a two-level system this is sufficient to infer the population of
$p_{22}$ but the presence of the third level means that $p_{11}(t)$ only
gives us limited information about the population of the other levels.

The existence of a non-zero detuning $\delta$ complicates the problem
substantially.  In the absence of anharmonicity, i.e., for $\delta=0$
analytic expressions for $p_{11}(t)$ can be obtained and the problem
reduced to a single frequency estimation problem~\cite{SL2010}.  For
$\delta\neq 0$ there is no simple closed form for the signal $p_{11}(t)$
and the eigenvalues are no longer of the form $0,\pm \lambda$, but are
instead $\omega_{12}=\lambda_2-\lambda_1=\omega-\Delta\omega$ and
$\omega_{23}=\lambda_3-\lambda_2 = \omega+\Delta\omega$.  For small
detunings $\delta$ relative to the coupling strengths $d_n$, the
frequency splitting $\Delta\omega$ is much smaller than $\omega$.  To
obtain obtain a frequency resolution of $\Delta\omega$ using spectral
analysis would require a signal length of at least $1/(\Delta\omega)$.
However, using the structure of the signal and restricting to solutions
consistent with our prior knowledge we can do considerably better.  We
know that the observed signal must be of the form
\begin{equation}
  p_{11}(t) = a_0 + a_1 \cos((\omega-\Delta\omega)t)
                  + a_2 \cos((\omega+\Delta\omega)t)
                  + a_3 \cos(2\omega t).
\end{equation}

A natural starting point for a maximum likelihood estimation is thus to
choose basis functions $g_0=1$, $g_1(t)=\cos((\omega-\Delta \omega)t)$,
$g_2(t)=\cos((\omega+\Delta \omega)t)$ and $g_3(t)=\cos(2\omega t)$, and
following standard techniques, maximize the log-likelihood
function~\cite{SO2009,Bretthorst88}
\begin{equation} 
  \label{eq:loglikelihood}
  L(\{\omega,\Delta\omega\}|\vec{d}) 
   \propto \frac{m_b-N_t}{2} \log_{10}
    \left[1-\frac{m_b \ave{\vec{h}^2}} {N_t\ave{\vec{d}^2}}\right],
\end{equation}
where $m_b=4$ is the number of basis functions, $N_t$ is the number of
data points and
\begin{equation}
 \ave{\vec{d}^2} = \frac{1}{N_t} \sum_{n=0}^{N_t-1} d_n^2, \quad
 \ave{\vec{h}^2} = \frac{1}{m_b} \sum_{m=0}^{m_b-1} h_m^2.
\end{equation}
The elements $h_m$ of the $(m_b,1)$-vector $\vec{h}$ are projections
of the $(1,N_t)$-data vector $\vec{d}$ onto a set of orthonormal basis
vectors derived from the non-orthogonal basis functions $g_m(t)$
evaluated at the respective sample times $t_n$.  Concretely, setting
$G_{mn}=g_m(t_n)$, let $\lambda_m$ and $\vec{e}_m$ be the eigenvalues
and corresponding (normalized) eigenvectors of the $m_b\times m_b$
matrix $G G^\dag$ with $G=(G_{mn})$, and let $E=(e_{m'm})$ be a matrix
whose columns are $\vec{e}_m$.  Then we have $H=V G$ and $\vec{h}=H
\vec{d}^\dag$ with $V =\diag(\alpha_m^{-1/2}) E^\dag$, and the
corresponding coefficient vector is $\vec{a}=\vec{h}^\dag
V$~\cite{SO2009}.

\begin{figure*}
\includegraphics[width=\textwidth]{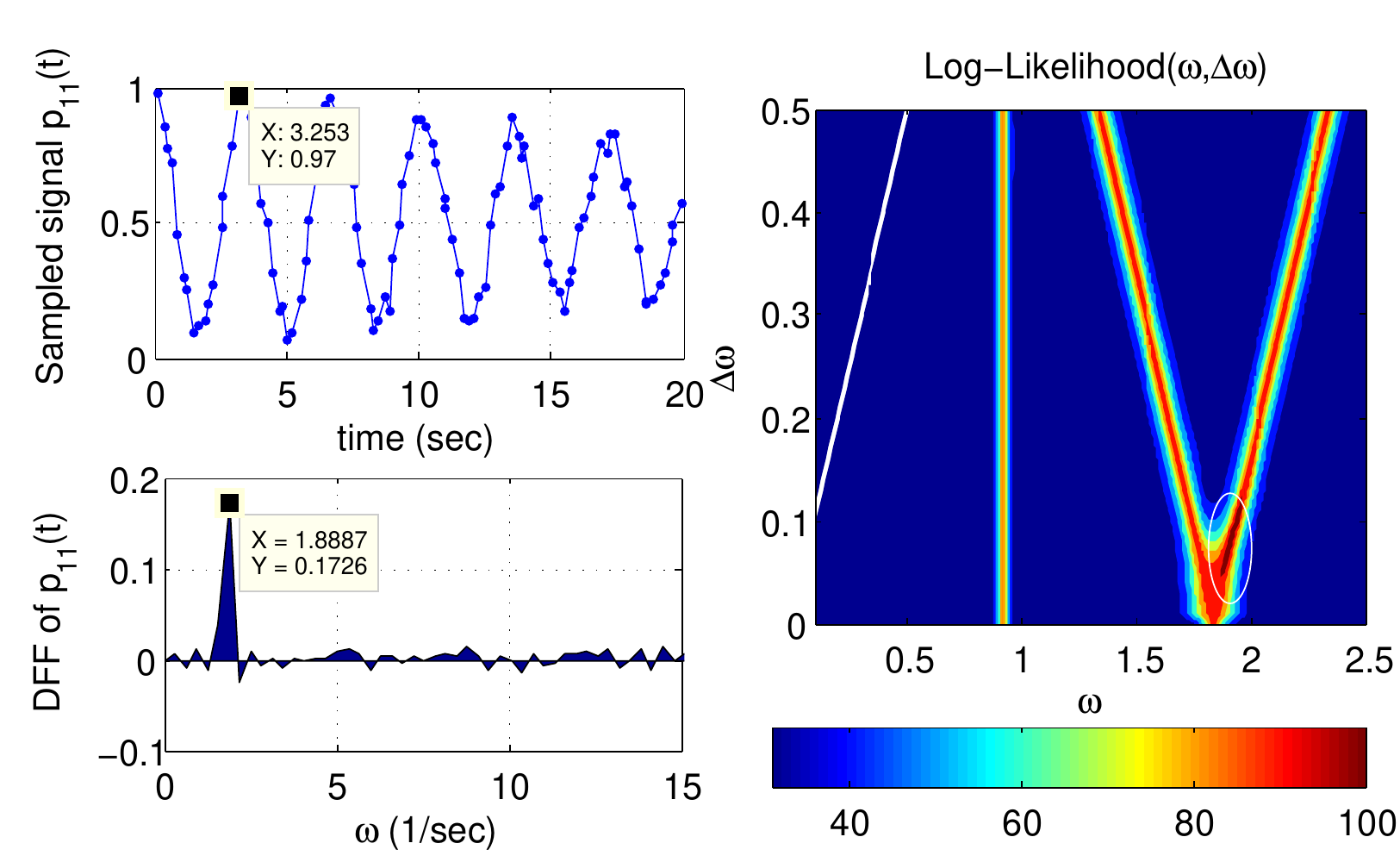}
\caption{``Qubit'' with leakage to a slightly detuned third level.
  The raw signal $p_{11}(t)$ (low-descrepancy sampling, $N_t=100$,
  $N_e=100$) provides (top, left) a preliminary estimate of the main
  frequency. The DFT (bottom left) shows only a single peak.  The
  log-likelihood (right) clearly shows that a model with a small
  splitting of $\Delta\omega$ around $0.1$ (white ellipse) is
  significantly more likely than a single frequency model.  (Colour
  online)} \label{fig1}
\end{figure*}

Fig.~\ref{fig1} shows that the log-likelihood function of the data
provides strong evidence for a non-zero detuning, even though no peak
splitting is detectable in the Fourier spectrum of the signal.  For the
given input data the log-likelihood function has a squeezed peak that is
narrow in one direction but much broader in the other.  The fact that
the peak is squeezed in a direction not aligned with a coordinate axis
shows that the uncertainties in $\omega$ and $\Delta\omega$ are not
independent.  The plot also shows that the width of the peak along the
$\Delta\omega$ direction is much greater than that in $\omega$ direction.

\begin{figure*}
\includegraphics[width=0.5\textwidth]{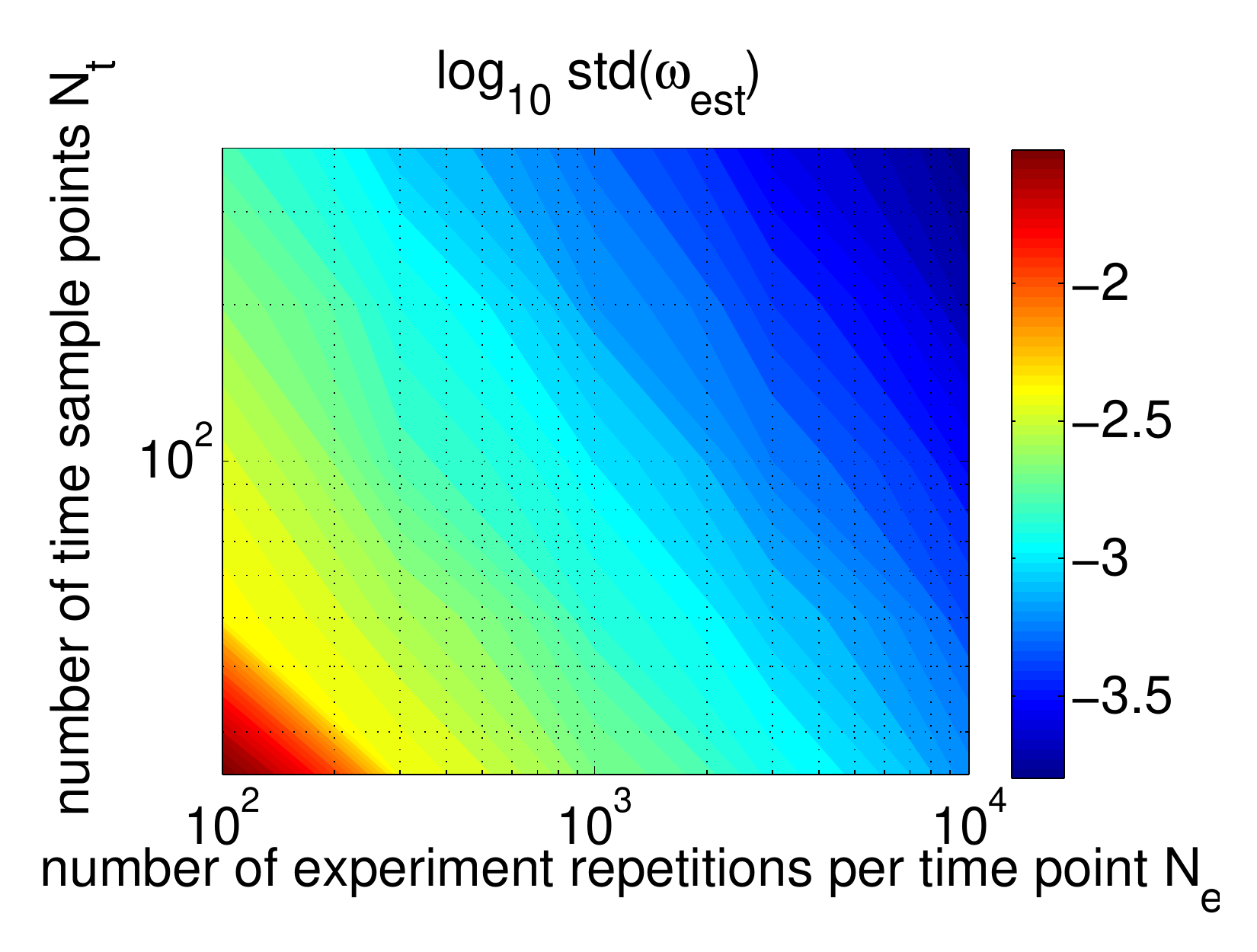}
\includegraphics[width=0.5\textwidth]{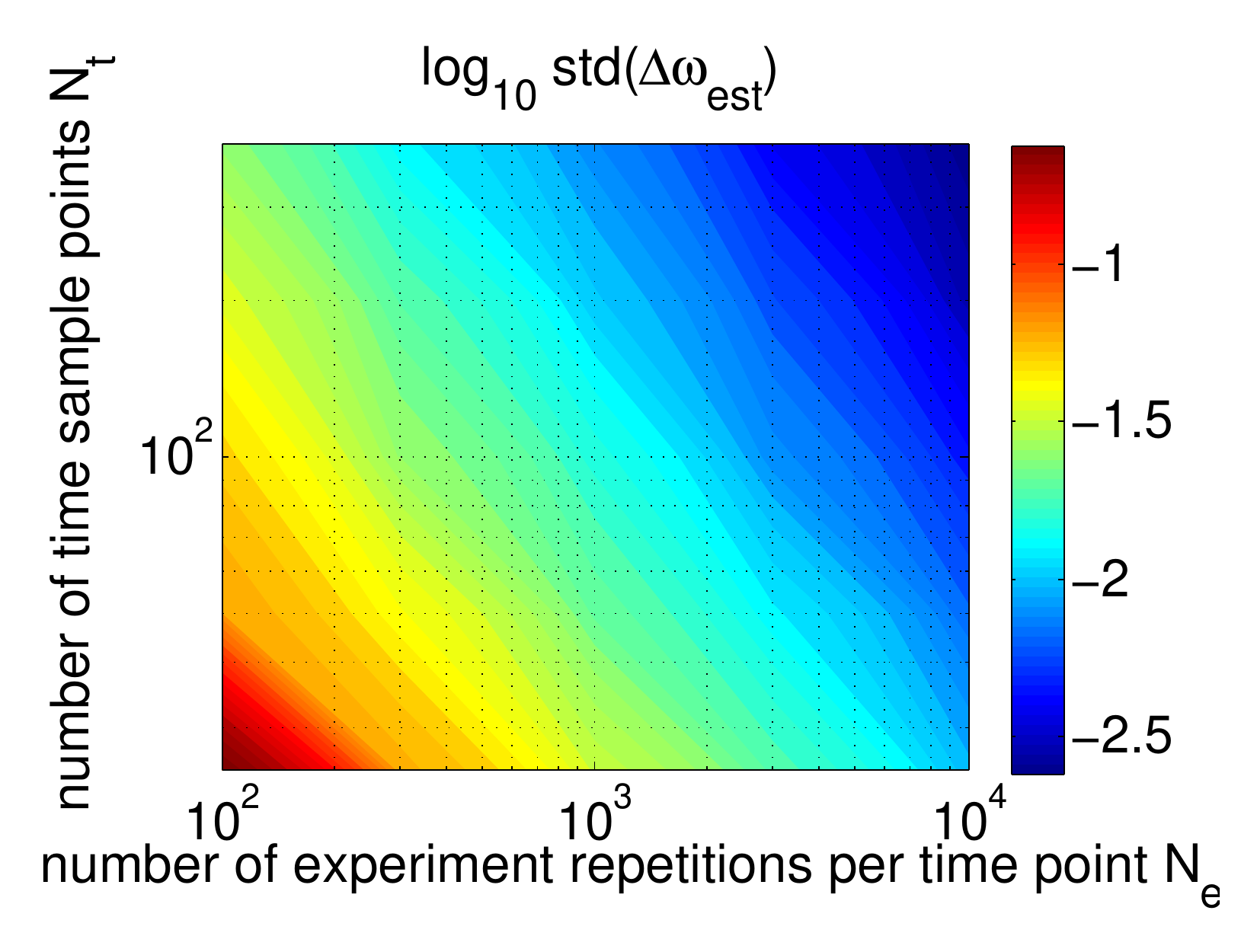}
\caption{Standard deviation of $\omega_{est}$ and $\Delta\omega_{est}$
  for 256 simulated experiments each as a function of the number of
  time samples $N_t$ and the number of experiment repetitions
  $N_e$. The scale is $\log_{10}$. (Colour online)}
\label{fig:uncert}
\end{figure*}

This is also reflected in the observed uncertainties of the estimates
of $\omega$ and $\Delta \omega$.  If we take the $(\omega_{est},\Delta
\omega_{est})$ to be the coordinates for which the log-likelihood
peaks then their standard deviations give an indication of the
uncertainty in our estimates.  Fig.~\ref{fig:uncert} shows the
standard deviation of $\omega_{est}$ and $\Delta\omega_{est}$ for 256
simulated experiments each as a function of the number of time samples
$N_t$ and the number of experiment repetitions $N_e$, on a logarithmic
scale.  The plots look qualitatively similar, suggesting a similar
scaling, but the scale shows that the uncertainty of the
$\Delta\omega$ estimates is about one order of magnitude greater than
that of the $\omega$ estimates.  We observe a similar scaling for the
estimated amplitudes $a_m$ for $m=0,1,2,3$ (not shown).  The median
relative error in the estimated Hamiltonian shows a similar behaviour
though with some kinks, and it is interesting to note that the
relative errors in the Hamiltonian tend to be larger than the errors
in the estimated frequencies and amplitudes of the signal.

\begin{figure*}
\includegraphics[width=0.5\textwidth]{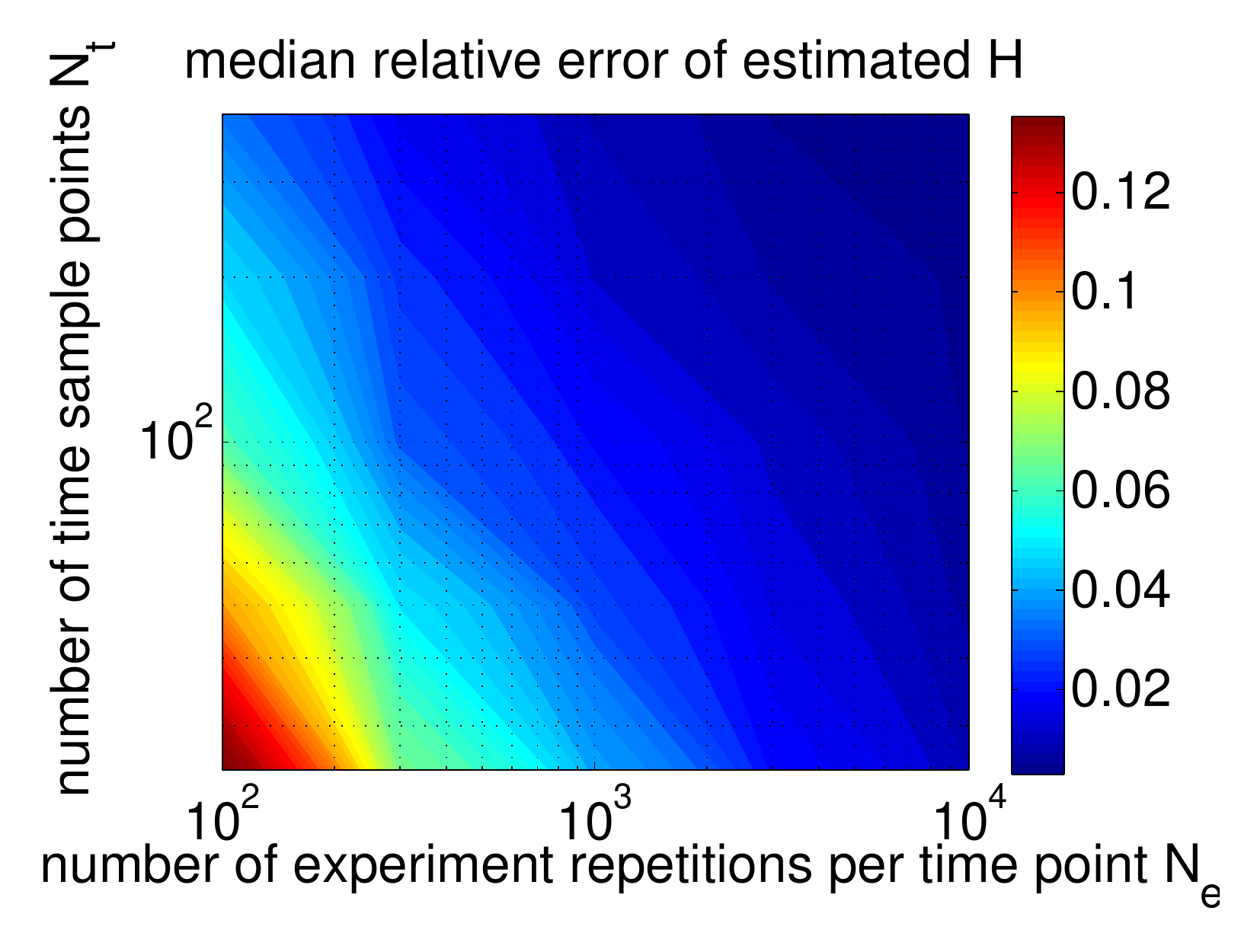}
\includegraphics[width=0.5\textwidth]{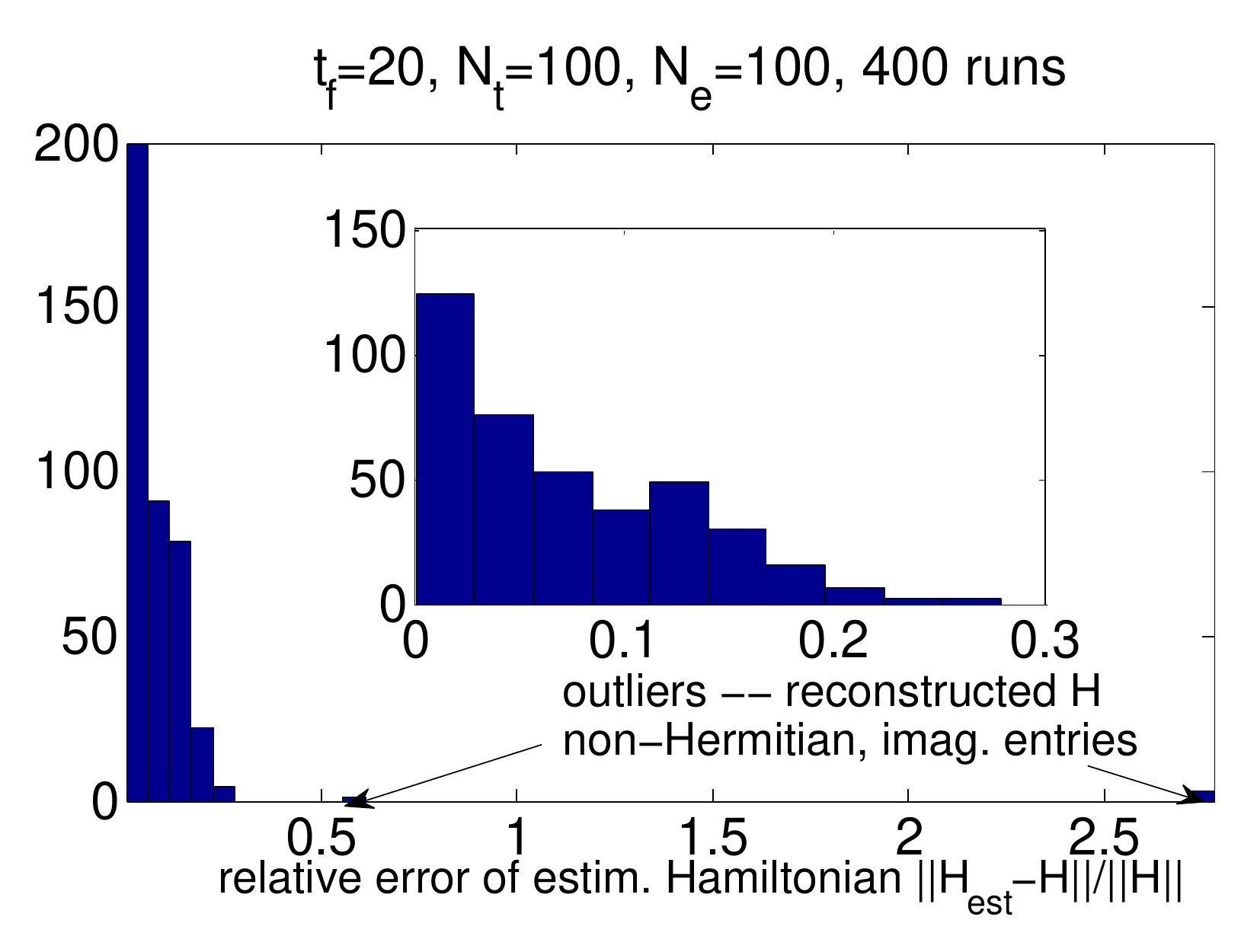}
\caption{(left) Error scaling as a function of sampling times and
  accuracy. (right) Histogram of reconstruction error showing outliers
  corresponding to unphysical solutions. (Colour online)}
\label{fig:error-H}
\end{figure*}

Preliminary results suggest that fragility of the reconstruction
procedure is responsible for the observed larger spread in the errors of
the reconstructed Hamiltonian even if the uncertainty of the estimated
signal parameters is quite low as shown in Fig.~\ref{fig:error-H}
(left).  The reconstruction procedure can sometimes fail, leading to
outliers in the relative Hamiltonian error histogram shown in
Fig.~\ref{fig:error-H} (right), which typically correspond to unphysical
Hamiltonians.

If the likelihood function does not have a sufficiently sharp peak, we
can adaptively refine the sampling to reduce the uncertainty. A simple
adaptive strategy is as follows:
\begin{enumerate}

\item Preliminary sampling: Over a pre-chosen sampling period, measure
  at randomly chosen sampling times and compute the likelihood of
  various models based on this initial data.

\item Uncertainty estimate: The uncertainty of the solution can be
  estimated from the sharpness and relative height of the highest
  peak in the likelihood plot.

\item Refinement: From the initial data, choose an ensemble of
  probable models and calculate the weighted expected variance of
  their data traces as a function of time.

\item New samples: We make additional measurements at times for which
  the most probable models differ the most and use the new data points
  to update the estimates of the signal parameters and Hamiltonian.

\item Repeat as necessary.

\end{enumerate}

Although similar in spirit to the adaptive Bayesian identification
strategy proposed Wiseman \emph{et al.} for single parameter
Hamiltonian estimation~\cite{SCCBW2011}, we do not minimize the
variance of a single parameter as the Hamiltonian depends on multiple
parameters. Another difficulty is that the multi-parameter likelihood
function is usually far from Gaussian and the expectation values of
$\omega$ and $\Delta\omega$ tend to differ substantially from the
maximum likelihood estimate. In this case using the expectation values
and variances with respect to a given parameter is not necessarily a
good indicator of the real uncertainty of the model.

\begin{figure}
\hspace{-0.5cm}
\includegraphics[width=0.5\textwidth]{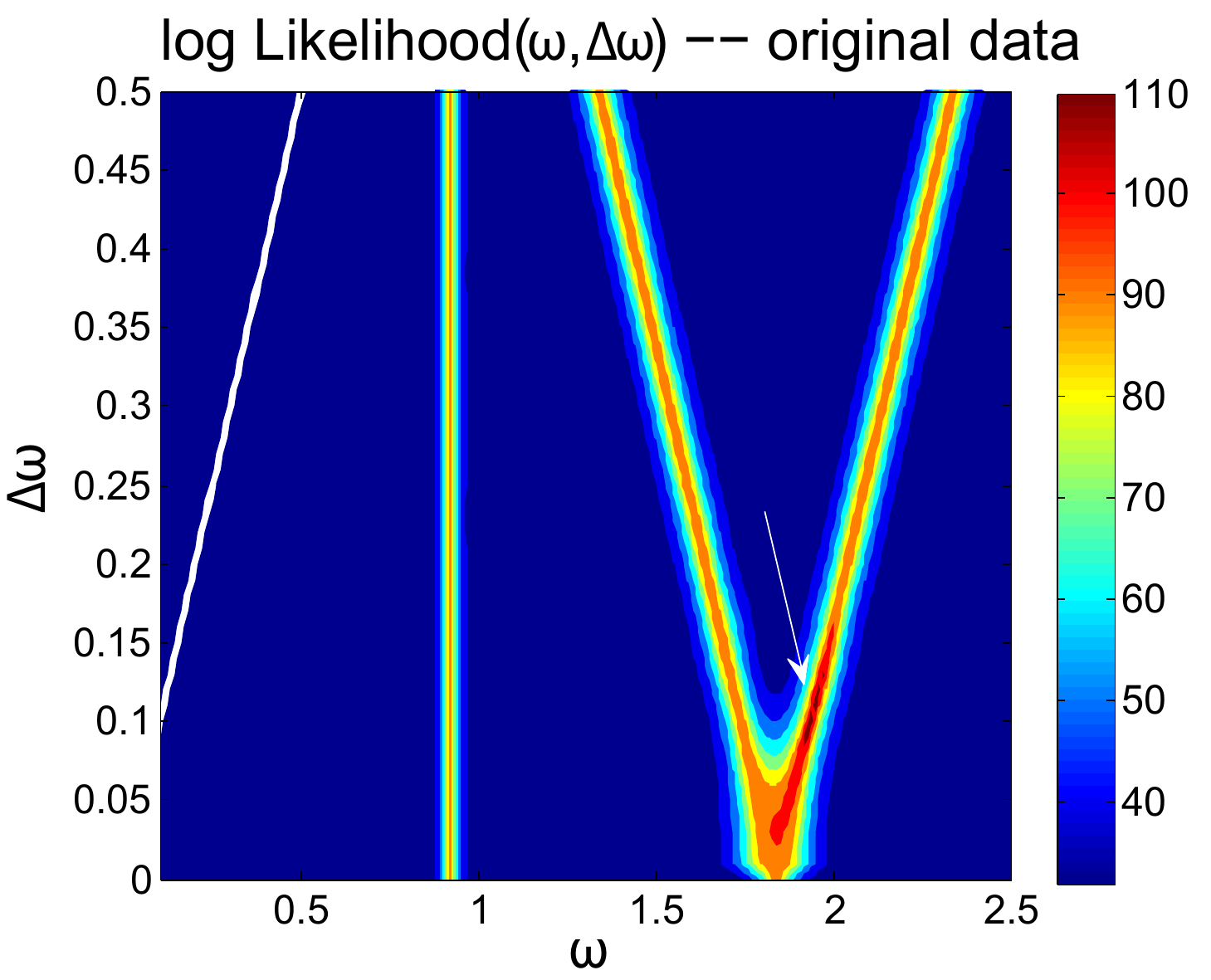}
\includegraphics[width=0.5\textwidth]{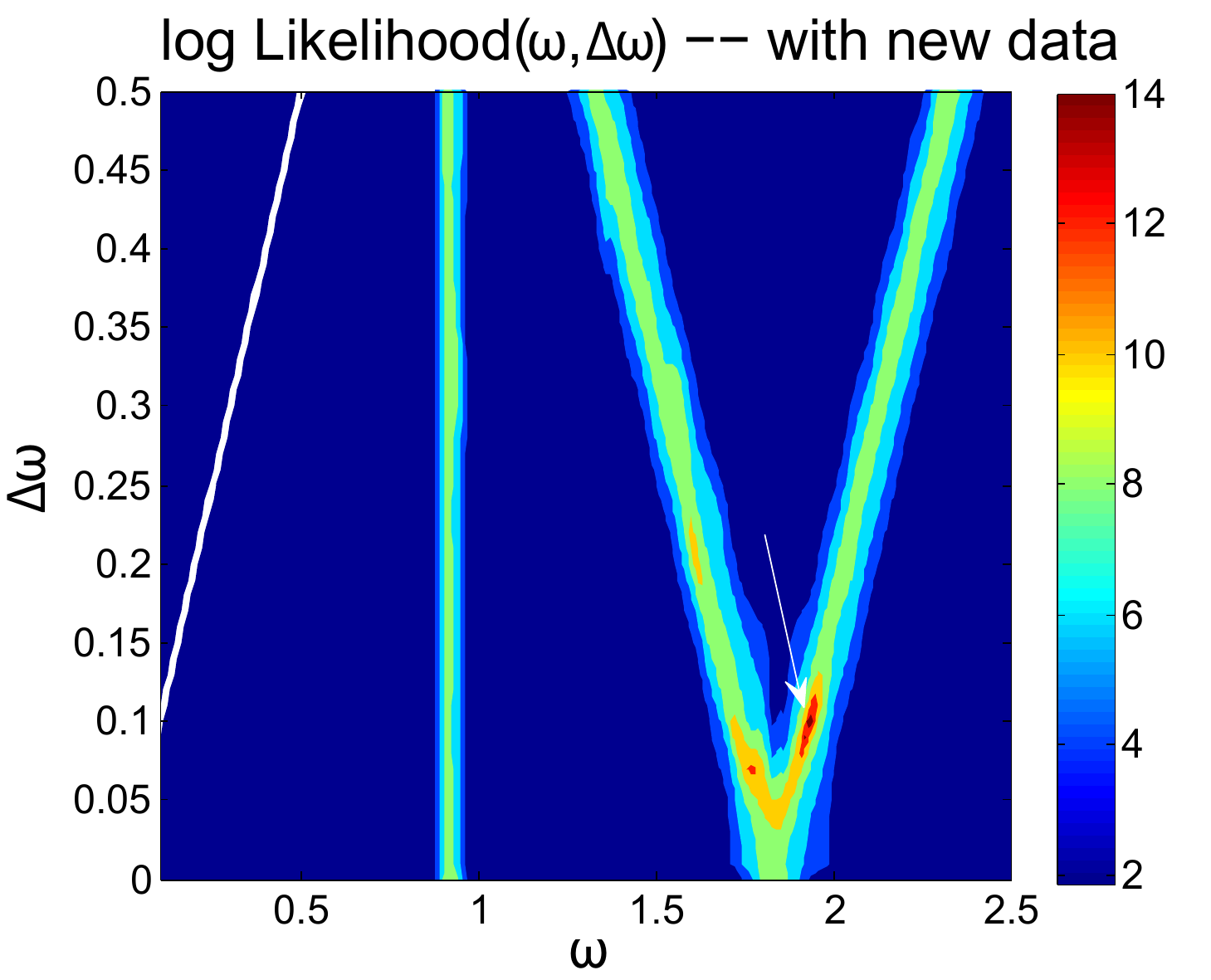}
\caption{Log-likelihood comparision (left) Initial, (right) After
  adaptive sampling. The probability is more highly concentrated and
  the accuracy of the solution greatly improved. (Colour
  online)} \label{fig:log-likelihood-compare}
\end{figure}

We apply this to the above qutrit system. Starting with a 100-point
low-discrepancy sampling of the selected time range $[0,20]$, we
obtain $N_e=100$ measurements per time point.  The resulting
log-likelihood function, shown in
Fig.~\ref{fig:log-likelihood-compare}(left), has a squeezed peak
centred at $(\omega,\Delta\omega)=(1.9468,0.1087)$.  To narrow the
peak width we resample using the initial estimate for
$(\omega,\Delta\omega)$. A simple and computationally cheap strategy
is to choose new sample points at integer multiples of
$T/2=\pi/\omega_{est}$, and get a few accurate samples using
e.g. $N_e=1000$. The idea is that we will be most sensitive to small
modulations in the peak heights due to $\Delta\omega$ at these times.
Indeed we find that the relative errors for the frequency and
amplitude estimates improve substantially,
Fig.~\ref{fig:log-likelihood-compare}(right). Yet, the relative error
of the reconstructed Hamiltonians does not always follow the same
trend and sometimes actually \emph{increases}. Further analysis
shows that this is due to the reconstruction step and the complex
dependence of $H$ on the estimated parameters.

This suggests that it would be desirable to estimate the Hamiltonian
paramaters directly from the data, maximizing
the likelihood
\begin{equation}
  \label{eq:P2}
  P(\vec{d}|\{\Omega,\alpha,\eps\}) \propto \exp \left[
	\sum_{j=1}^{N_t} |d_j -p_{11}(t_j)|^2)\right]^{-N_t/2}
\end{equation}
or its logarithm, and for convenience we have defined
$d_1=\Omega\cos{\alpha}$, $d_2=\Omega\sin{\alpha}$, and
$\delta=4\epsilon$. Although there is no simple closed form for
$p_{11}(t_j)$ in this case, it can be computed numerically and the
challenge is finding the global optimum of the 3-parameter likelihood
function. Without any prior information, we first compute the
log-likelihood on a 3D grid of parameter values, find the region where
the global optimum is expected, and use this information as a starting
point of a local optimization routine to find the peak. For the
example above, this yields
$(\Omega,\alpha,\eps)_{opt}=(1.7159,0.9494,0.5340)$ for the initial
data, versus $(\Omega,\alpha,\eps)_{opt}=(1.7323,0.9557,0.4999)$ with
the new data (actual $(1.7321,0.9553,0.5000)$).  The relative error of
the reconstructed Hamiltonian for the inital data is $0.0363$,
comparable (even slightly higher) to the estimate obtained using the
previous two-step approach. Using the new data, the relative error is
substantially lower, $3.8672 \times 10^{-4}$ if we maximize
(\ref{eq:P2}) instead of maximizing (\ref{eq:loglikelihood}) followed
by reconstruction.  This approach appears suitable for systematic
adaptive estimation, which will be investigated further in future
work.

\section{Conclusion}

We have outlined the problem of system characterisation and why it is an
essential basic building block of quantum control for many prospective
quantum information processing devices. We have
shown that such a bootstrapping procedure is possible and that much
information can be gained through utilisation of limited \emph{in situ}
initially present operational capabilities. By exploiting prior
knowledge and reasonable assumptions on the structure and behaviour of
the system, maximum likelihood analysis offers efficient
and robust characterisation and reconstruction of complex systems.

Scalability remains a challenge.  In the general setting, slightly
increasing the size of the system leads to an explosion in signal
complexity~\cite{SO2009} and directly applying system identification to
three of more qubits is a formidable task. Some
complexity reduction can be achieved using Bayesian signal
estimation in order to split up frequency and amplitude estimation,
though this has some drawbacks in ensuring physically allowed
reconstructed Hamiltonians. Extending the Bayesian estimation directly
to Hamiltonians would alleviate reconstruction validity but direct
optimisation of the likelihood is challenging for more than a few
parameters. Hence there is the need for a similar complexity reduction
for Hamiltonian parameter estimation. Exploiting as much structural
information about the system is essential in making the problem
tractable, especially in the case of restricted resources.

Adaptive estimation for multiple parameters is a ripe area for further
exploration. Practical online schemes may require pragmatic methods of
determining adaptive measurements, as full Bayesian optimisation
involves integration over a highly peaked payoff function in a high
dimensional parameter space. Development of effective yet
computationally efficient optimisation routines is imperative.

Relaxing more assumptions or expanding the set of resources available
would give greater experimental relevance. Preparation and measurement
capabilities may vary, for instance instead of projective measurement,
some experimental proposals implement continuous~\cite{ContMeas}, weak
or generalized measurements. States may also be prepared by relaxation
or adiabatic passage and may not coincide with the measurement basis.

More generally, there are interesting connections between the
compressive sensing (CS) and sparse reconstruction
paradigm~\cite{CSreference,CSGross} and how our model-based systen
characterisation techniques work. Instead of the union of
low-dimensional (linear) sub-spaces model in CS, we instead have a
solution space as the union of low-dimensional manifolds of
parameters. An extension of the notion of ``basis incoherence'' and
general techniques for efficient reconstruction from sparse data would
be extremely beneficial.

\end{document}